\documentclass[prb,twocolumn,amsfonts]{revtex4-1}
\usepackage{graphicx,graphics,color,epsfig}
\usepackage{bm}
\usepackage{amsmath}
\usepackage{amssymb}
\usepackage{epstopdf}
\begin{document}
\title{$U(1)$ Slave-spin theory and its application to Mott transition in a multi-orbital model for iron pnictides}
\author{Rong Yu}
\affiliation{Department of Physics \& Astronomy, Rice University, Houston, TX 77005}
\author{Qimiao Si}
\affiliation{Department of Physics \& Astronomy, Rice University, Houston, TX 77005}

\begin{abstract}
A $U(1)$ slave-spin representation is introduced for multi-orbital Hubbard models. As with the
$Z_2$ form
of L. de'Medici {\it et al.} (Ref.~\onlinecite{deMedici05}),
this approach represents a physical electron operator as the product of
a slave spin and an auxiliary fermion operator.
For non-degenerate multi-orbital models, our $U(1)$ approach is advantageous
in that it captures the non-interacting limit at the mean-field level.
For systems with either a single orbital or degenerate multiple orbitals,
the $U(1)$ and $Z_2$ slave-spin approachs
yield the same results in the
slave-spin-condensed phase.
In general, the $U(1)$ slave-spin approach contains a $U(1)$ gauge redundancy, and properly describes
a Mott insulating phase.
We apply the $U(1)$ slave-spin approach
to study the metal-to-insulator transition in a five-orbital model
for parent iron pnictides. We demonstrate a Mott transition as a function of the interactions in this model.
The nature of the Mott insulating state is influenced by the interplay between
the Hund's rule coupling and crystal field splittings. In the metallic phase,
when the Hund's rule coupling is beyond a threshold, there is
a crossover from a weakly correlated metal
to a strongly correlated one, through which the quasiparticle speactral
weight rapidly drops. The existence of such a strongly correlated metallic phase
supports the incipient Mott picture of the parent iron pnictides.
In the parameter regime for this phase and in the vicinity of the Mott transition,
we find that an orbital selective Mott state has nearly as competitive a 
ground state energy.
\end{abstract}
\maketitle

\section{Introduction}
Many important questions remain on the
physics of the iron pnictides and related
iron-based high-$T_c$
superconductors~\cite{Kamihara_FeAs,Zhao_CPL08,Rotter08,Hsuetal08,JGuo10,MFang11}.
One central issue is the strength of electron correlations these systems contain. The metallic nature and the collinear antiferromagnetic (AFM) ground state~\cite{delaCruz} in the parent pnictides
can arise within a weak-coupling approach, in which the Fermi surface nesting
plays an important role~\cite{Dongetal08}.
On the other hand, the large room-temperature electrical resistivity,
showing the bad metal behavior with the electron mean free path on the order of the inter-electron spacing,
as well as the suppression of Drude weight in optical conductivity~\cite{Qazilbash09,Huetal09,SiNatPhys09},
the well defined zone-boundary spin waves~\cite{ZhaoNatPhys09},
and the renormalization of LDA bandstructure in the ARPES measurements~\cite{Yietal09}
provide evidence for sufficiently strong electron correlations such that
 the system is in close proximity to a Mott transition
 with dominant contributions to the spin spectral weight from quasi-local
moments~\cite{SiAbrahams08,SiNJP09,Haule,Kutepov10,Yildirim,Ma,Fang:08,Xu:08,Daietal09,Uhrig,
KSeo,WChen,PGoswami,Moreo,Berg}.

This incipient Mott picture is further supported by the properties of iron
chalcogenides~\cite{Hsuetal08,JGuo10,MFang11}.
In either the 11- or the 122-chalcogenides, the magnetic ordering wave vector and the large magnetic
moment~\cite{Lipscombe11,Bao122A,Bao122B,Ye122} (ranging from $2$ to $3.4$ $\mu_B$) can hardly
be explained by a Fermi surface nesting mechanism, but is readily understood within a quasi-local moment
model~\cite{MaFeTe,HuJ1J2J3,CaoDai11,YuKFeSeMag}.
It has also been shown that the band narrowing effect, either due to the expansion of the iron lattice unit cell in a iron oxychalcogenide~\cite{Zhuetal10}, or from the ordered iron vacancies in 122 iron selenides~\cite{CaoDai11,YuVacOrder11,Zhou11}, may drive the system through the Mott transition to a Mott insulator.

In general, the degree of electron correlations can be measured by the ratio $U/D$, where $U$ refers to a characteristic interaction strength, such as the Coulomb repulsion in a Hubbard model, and $D$ is the full bandwidth, a scale of the kinetic energy of the system. A Mott transition separating the metallic and the Mott insulating phases takes place at $U_c\sim D$. In the metallic phase, electron correlations can alternatively be measured by the quasiparticle spectral weight $Z$, which is unity at the non-interacting limit $U=0$, and vanishes at the Mott transition $U=U_c$. The incipient Mott picture relies on the existence of a (putative) Mott transition, and assume that the system is not too far from this transition so that $Z$ is relatively small.

The above considerations suggest that it is very important to theoretically investigate
how a metal to Mott insulator transition takes place in a model which
is applicable to the parent iron pnictides and/or chalcogenides. In a previous paper,
by studying a two-orbital and a four-orbital model, the authors have shown
that a Mott transition generally exists in these models~\cite{Yu11}.
But both models are at half-filling. By contrast, in the parent iron pnictides, six electrons occupy five $3d$ orbitals of each Fe. This means that the system is away from half-filling. It would be important to investigate extent to which the Mott insulating states persist in such a situation.

Historically, the Mott transition in a single-orbital Hubbard model has been
studied by using various techniques, such as Dynamical mean-field theory (DMFT)~\cite{}
and Gutzwiller approximation~\cite{}. Among these methods, the Kotliar-Ruckenstein
slave-boson method~\cite{KotliarRuckenstein} has been broadly used. In this theory,
a Mott transition at finite $U/D$ is obtained already at the mean-field level. However,
it is difficult to apply this approach to multi-orbital systems as it would introduce $4^M$
slave boson fields for a model with $M$ orbitals;
the number of variational
variables is already huge even for $M=5$. Recently several other slave particle theories following the idea
of charge-spin separation have been proposed. These include the slave-rotor
 theory~\cite{FlorensGeorges}, and the slave-spin theory~\cite{deMedici05}.
 In both theories, a slave bosonic variable
  (a quantum $O(2)$ rotor in the slave-rotor theory, and a quantum $S=1/2$ spin in the slave-spin theory) is introduced to carry the electric charge, and an auxiliary fermion (the spinon) carries the spin of the electron.
 The metallic phase corresponds to the state that the slave particles are Bose condensed, so that charge excitations
 are gapless along with the spin excitations.
By contrast, the Mott insulator corresponds to the state that the slave particles are disordered and gapped;
charge excitations are gapped while the spin excitations remain gapless.
The slave rotor method is very economical because it introduces only one rotor per site taking
account of the total charge. It is also very efficient if the interactions have an $SU(2M)$ symmetry for an
 $M$-orbital system. However, it can not be easily applied to systems with a non-zero Hund's rule coupling
 which breaks the $SU(2M)$ symmetry. It is also not convenient to handle systems that exhibit
 strong orbital dependence, such as the orbital selective Mott transition (OSMT). The slave-spin theory
overcomes these drawbacks by introducing a slave spin for each orbital and spin flavor.
Compared with the Kotliar-Ruckenstein slave-boson theory, it is still very economical because it introduces
only $2M$ slave spins per site. It has been successfully used to study the Mott transition and
OSMT in multi-orbital systems with a non-zero Hund's rule coupling~\cite{deMedici11}.
There are however several issues with the slave-spin representation.
 First, it has difficulties when applied to a multi-orbital system in which one or more orbitals are away from half-filling
due to crystal field splitting. Secondly, in its original construction,
the slave-spin representation has a $Z_2$ gauge redundancy.
This makes the spinons carry both spin and charge currents, and causes difficulties in describing a Mott insulating phase ~\cite{NandkishoreSenthil12}.

In this paper we propose an
$XY$ slave-spin theory that
is free of these issues.
This slave-spin theory has a $U(1)$ gauge redundancy, and properly describes Mott insulating phases.
We develop a mean-field theory, which can be applied to multi-orbital systems with a non-zero crystal field splitting.
For a model with a single orbital or degnerate multiorbitals, on the other hand,
our $U(1)$ slave-spin mean-field theory and the $Z_2$ slave-spin theory~\cite{Sigrist} give the same results in their spin-condensed phases.
We then apply our formulation to study the Mott transition in a five-orbital model for the parent iron pnictides.
We establish the existence of a Mott transition in this model.
Both the nature of the metallic and Mott insulating phases are strongly affected by the interplay
of Hund's rule coupling and the crystal field splitting. A crossover to a strongly correlated metallic state exists
when the Hund's rule coupling is beyond a threshold. The existence of this state is in agreement with the
 incipient Mott picture.

The rest of the paper is organized as follows. In Sec.~\ref{Sec:Method} we first introduce our construction of the $U(1)$ slave-spin theory, and develop a mean-field theory based on this new construction. We also compare our construction with the slave-rotor and $Z_2$ slave-spin theories.
In Sec.~\ref{Sec:MottTransition} we apply the $U(1)$ slave-spin mean-field theory to study the Mott transition in a five-orbital model for the parent iron pnictides, and show how the transition is affected by the interplay
between Hund's rule coupling and the crystal field splitting. Finally Sec.~\ref{Sec:Conclusions} contains
some concluding remarks.

\section{Method}\label{Sec:Method}
In this paper we are interested in the metal-to-insulator transition in a multi-orbital system, for which the Hamiltonian reads
\begin{equation}
 \label{Eq:Ham_tot} H=H_0 + H_{\mathrm{int}}.
\end{equation}
$H_0$ contains the tight-binding parameters among the multiple orbitals,
\begin{equation}
 \label{Eq:Ham_0} H_0=\frac{1}{2}\sum_{ij\alpha\beta\sigma} t^{\alpha\beta}_{ij} d^\dagger_{i\alpha\sigma} d_{j\beta\sigma} + \sum_{i\alpha\sigma} (\Delta_\alpha-\mu) d^\dagger_{i\alpha\sigma} d_{i\alpha\sigma},
\end{equation}
where $d^\dagger_{i\alpha\sigma}$ creates an electron in orbital $\alpha$ with spin $\sigma$ at site $i$, $\Delta_\alpha$
is the on-site energy reflecting the crystal field splitting, and $\mu$ is the chemical potential. $H_{\rm{int}}$
contains on-site Hubbard interactions
\begin{eqnarray}
 \label{Eq:Ham_int} H_{\rm{int}} &=& \frac{U}{2} \sum_{i,\alpha,\sigma}n_{i\alpha\sigma}n_{i\alpha\bar{\sigma}}\nonumber\\
 &&+\sum_{i,\alpha<\beta,\sigma} \left\{ U^\prime n_{i\alpha\sigma} n_{i\beta\bar{\sigma}}\right. 
 + (U^\prime-J) n_{i\alpha\sigma} n_{i\beta\sigma}\nonumber\\
&&\left.-J(d^\dagger_{i\alpha\sigma}d_{i\alpha\bar{\sigma}} d^\dagger_{i\beta\bar{\sigma}}d_{i\beta\sigma}
 -d^\dagger_{i\alpha\sigma}d^\dagger_{i\alpha\bar{\sigma}}
 d_{i\beta\sigma}d_{i\beta\bar{\sigma}}) \right\}.
\end{eqnarray}
where $n_{i\alpha\sigma}=d^\dagger_{i\alpha\sigma} d_{i\alpha\sigma}$. In this model,
$U$, $U^\prime$, and $J$ respectively denote the intraorbital repulsion, the interorbital repulsion,
and the Hund's rule exchange coulping.
In the following, we will take $U^\prime=U-2J$.~\cite{Castellani78}

\subsection{$U(1)$ slave-spin theory}
To study this multi-orbital model, we implement the idea of charge and spin separation of the
$d$ electrons by using the slave-spin approach. For each orbital and spin flavor we
rewrite the electron creation operator to be
\begin{equation}
 \label{Eq:SSCreate} d^\dagger_{i\alpha\sigma} = S^+_{i\alpha\sigma} f^\dagger_{i\alpha\sigma},
\end{equation}
where $S^+_{i\alpha\sigma}$ is a ladder operator of the slave quantum $S=1/2$
spin carrying the charge degree of freedom of the electron, and $f^\dagger_{i\alpha\sigma}$ is a
spinon creation operator. We further enforce a constraint for each site
\begin{equation}
 \label{Eq:constraint} S^z_{i\alpha\sigma} = f^\dagger_{i\alpha\sigma} f_{i\alpha\sigma} - \frac{1}{2} ,
\end{equation}
which  restricts  the Hilbert space to the physical one.

Note that our slave-spin formulation in Eq.~\eqref{Eq:SSCreate} is different from
that introduced in Refs.~\onlinecite{deMedici05,HassandeMedici10}, in which
\begin{equation}\label{Eq:OSSCreate}
d^\dagger_{i\alpha\sigma} = O^\dagger_{i\alpha\sigma} f^\dagger_{i\alpha\sigma},
\end{equation}
where
\begin{equation}\label{Eq:OSS}
O^\dagger_{i\alpha\sigma}=S^+_{i\alpha\sigma} + c_{i\alpha\sigma} S^-_{i\alpha\sigma},
\end{equation}
with $c_{i\alpha\sigma}$ being a complex number.
In that formulation, the gauge redundancy is reduced from $U(1)$ to $Z_2$ in Eq.~\eqref{Eq:OSSCreate} and
Eq.~\eqref{Eq:OSS} due to the mixing between $S^+_{i\alpha\sigma}$ and $S^-_{i\alpha\sigma}$
(hence they are referred to as a $Z_2$ slave-spin representation);
as a consequence, the slave spins can not carry the $U(1)$ charge.

In our Eqs.~\eqref{Eq:SSCreate},\eqref{Eq:constraint}, there is a $U(1)$ gauge redundancy corresponding to
$f^\dagger_{i\alpha\sigma}\rightarrow f^\dagger_{i\alpha\sigma} e^{-i\theta_{i\alpha\sigma}}$
and $S^+_{i\alpha\sigma}\rightarrow S^+_{i\alpha\sigma} e^{i\theta_{i\alpha\sigma}}$.
The slave spins carry the $U(1)$ charge, similarly as the slave rotors ~\cite{FlorensGeorges}.
We will refer to this as a $U(1)$ slave-spin theory. In our construction, the phase that
corresponds to disordered slave spins
(preserving the $U(1)$ symmetry) but with gapless spinons corresponds to a Mott insulator.

Next we develop a mean-field theory based on the construction of
Eq.~\eqref{Eq:SSCreate} and Eq.~\eqref{Eq:constraint}. A naive mean-field theory based on
Eq.~\eqref{Eq:SSCreate} would not produce the correct
quasiparticle spectral weight in the non-interacting
limit~\cite{HassandeMedici10}.
To make progress, we rewrite the slave spin operators in their Schwinger boson representation:
$S^+_{i\alpha\sigma} = a^\dagger_{i\alpha\sigma} b_{i\alpha\sigma}$, $S^-_{i\alpha\sigma}
= b^\dagger_{i\alpha\sigma} a_{i\alpha\sigma}$, and $S^z_{i\alpha\sigma} = (a^\dagger_{i\alpha\sigma}
a_{i\alpha\sigma} - b^\dagger_{i\alpha\sigma} b_{i\alpha\sigma})/2$. The constraint in Eq.~\eqref{Eq:constraint}
then becomes $a^\dagger_{i\alpha\sigma} a_{i\alpha\sigma} - b^\dagger_{i\alpha\sigma} b_{i\alpha\sigma}
= 2 f^\dagger_{i\alpha\sigma} f_{i\alpha\sigma} -1$. Here we need to introduce an extra constraint
$a^\dagger_{i\alpha\sigma} a_{i\alpha\sigma} + b^\dagger_{i\alpha\sigma} b_{i\alpha\sigma} = 1$
so that the Schwinger bosons represent $S=1/2$ spins. We then see that $a_{i\alpha\sigma}$
and $b_{i\alpha\sigma}$ are hard-core bosons. In light of the Kotliar-Ruckenstein
slave-boson mean-field theory~\cite{KotliarRuckenstein}, we now define a dressed operator in the Schwinger
boson representation which automatically takes account this constraint:
\begin{equation}
 \label{Eq:Zdagger} z^\dagger_{i\alpha\sigma} = P^+_{i\alpha\sigma} a^\dagger_{i\alpha\sigma} b_{i\alpha\sigma}
 P^-_{i\alpha\sigma},
\end{equation}
where $P^\pm_{i\alpha\sigma}=1/\sqrt{1/2+\delta \pm (a^\dagger_{i\alpha\sigma} a_{i\alpha\sigma}
- b^\dagger_{i\alpha\sigma} b_{i\alpha\sigma})/2}$, and $\delta$ is an infinitesimal positive
number to regulate $P^\pm_{i\alpha\sigma}$.
$z^\dagger$ and $a^\dagger b$ are equivalent in the physical Hilbert space. In the Schwinger boson
representation, Eq.~\eqref{Eq:SSCreate} becomes \begin{equation}\label{Eq:SBcreate}
d^\dagger_{i\alpha\sigma}=z^\dagger_{i\alpha\sigma} f^\dagger_{i\alpha\sigma}.
\end{equation}

At the mean-field level, we treat the constraint Eq.~\eqref{Eq:constraint} on average by introducing
a Lagrange multiplier, and decompose the boson and spinon operators. We obtain
two mean-field Hamiltonians respectively for the spinons and the Schwinger bosons:
\begin{eqnarray}
 \label{Eq:Hf}  H^{\mathrm{mf}}_f &=& \frac{1}{2}\sum_{ij\alpha\beta\sigma} t^{\alpha\beta}_{ij}
 \langle z^\dagger_{i\alpha\sigma} z_{j\beta\sigma}\rangle f^\dagger_{i\alpha\sigma} f_{j\beta\sigma}
 \nonumber\\
 && + \sum_{i\alpha\sigma}  (\Delta_\alpha - \lambda_{i\alpha\sigma}-\mu) f^\dagger_{i\alpha\sigma}
 f_{i\alpha\sigma}  ,\\
 \label{Eq:HS}  H^{\mathrm{mf}}_S &=& \frac{1}{2}\sum_{ij\alpha\beta\sigma} t^{\alpha\beta}_{ij}
 \langle f^\dagger_{i\alpha\sigma} f_{j\beta\sigma} \rangle z^\dagger_{i\alpha\sigma} z_{j\beta\sigma}\nonumber\\
 && + \sum_{i\alpha\sigma} \frac{\lambda_{i\alpha\sigma}}{2} (\hat{n}^a_{i\alpha\sigma}
 - \hat{n}^b_{i\alpha\sigma}) + H^S_{\mathrm{int}},
\end{eqnarray}
where $\langle\cdots\rangle$ denotes the mean-field value,
$\hat{n}^{a}_{i\alpha\sigma}=a^\dagger_{i\alpha\sigma} a_{i\alpha\sigma}$,
and $\lambda_{i\alpha\sigma}$ is the Lagrange multiplier to handle the constraint in Eq.~\eqref{Eq:constraint}.
The quasiparticle spectral weight is defined as $Z_{i\alpha\sigma}=|\langle z_{i\alpha\sigma}\rangle|^2$. In Eq.~\eqref{Eq:HS} $H^S_{\mathrm{int}}$ refers to the interaction Hamiltonian in the Schwinger boson representation. It can be obtained by rewriting Eq.~\eqref{Eq:Ham_int} in the slave-spin representation $H_{\mathrm{int}}\rightarrow H_{\mathrm{int}}(\mathbf{S})$,\cite{Yu11} then substitute the Schwinger bosons for the spin operators.
The mean-field Hamiltonian $H^{\mathrm{mf}}_S$ has an internal $U(1)$ symmetry of the bosons.
For a single orbital, it is a Bose Hubbard model for two species of bosons, and is equivalent to a model of interacting XY spins in a magnetic field.
At commensurate fillings, by breaking the internal $U(1)$ symmetry, this model has a phase
transition from a bosonic Mott insulator to a superfluid with decreasing the interactions. These two
phases correspond to the Mott insulating and metallic states in the original $d$ electron problem.
We then approach the Mott transition from the ordered phase, and further adopt a single-site approximation
to Eq.~\eqref{Eq:Hf} and Eq.~\eqref{Eq:HS} with the decoupling $z^\dagger_{i\alpha\sigma}
z_{j\beta\sigma} \approx \langle z^\dagger_{i\alpha\sigma}\rangle z_{j\beta\sigma}
+  z^\dagger_{i\alpha\sigma} \langle z_{j\beta\sigma}\rangle - \langle
z^\dagger_{i\alpha\sigma}\rangle \langle z_{j\beta\sigma}\rangle$.
For simplicity, we focus on the paramagnetic phase, and assume translational symmetry.
These allow us to drop the spin and site indices in the formulas. The mean-field boson Hamiltonian then reads
\begin{eqnarray}
\label{Eq:HSsinglesite} H^{\mathrm{mf}}_S &\approx& \sum_{\alpha\beta} \epsilon^{\alpha\beta}
\left(\langle z^\dagger_{\alpha}\rangle z_{\beta} + \langle z_{\beta}\rangle z^\dagger_{\alpha} \right) \nonumber\\
&& + \sum_\alpha \frac{\lambda_{\alpha}}{2} (\hat{n}^a_{\alpha} - \hat{n}^b_{\alpha}) + H^S_{\mathrm{int}},
\end{eqnarray}
where $\epsilon^{\alpha\beta} = \sum_{ij\sigma} t^{\alpha\beta}_{ij} \langle f^\dagger_{i\alpha\sigma}
f_{j\beta\sigma}\rangle/2$.
In Eq.~\eqref{Eq:HSsinglesite}, we Taylor-expand $z_\alpha$ and $z^\dagger_\alpha$
in terms of $\hat{A}-\langle \hat{A}\rangle$ (where $\hat{A}=\hat{n}^a, \hat{n}^b, a^\dagger b$), and keep
up to the linear terms in $\hat{A}-\langle \hat{A}\rangle$, obtaining
\begin{equation}
 \label{Eq:Zapprox} z^\dagger_\alpha \approx \tilde{z}^\dagger_\alpha
 + \langle\tilde{z}^\dagger_\alpha\rangle \eta_\alpha [\hat{n}^a_\alpha-\hat{n}^b_\alpha-(2n^f_\alpha-1)],
\end{equation}
where $\tilde{z}^\dagger_\alpha = \langle P^+_\alpha\rangle a^\dagger_\alpha b_\alpha \langle
P^-_\alpha\rangle$, $\eta_\alpha = (2n^f_\alpha-1)/[4n^f_\alpha(1-n^f_\alpha)]$, $n^f_\alpha
=\frac{1}{N}\sum_k\langle f^\dagger_{k\alpha} f_{k\alpha}\rangle$, and $n^f_\alpha =
\langle \hat{n}^a_\alpha\rangle=1-\langle \hat{n}^b_\alpha\rangle$ from the constraints.
We find that Eq.~\eqref{Eq:Zapprox} already gives good mean-field results, and will hence
drop the higher order terms in the expansion. Note that the approximate form
of $z^\dagger_\alpha$ in Eq.~\eqref{Eq:Zapprox} is only used to simplify $H^{\mathrm{mf}}_S$,
but can not be fed into Eq.~\eqref{Eq:SBcreate} to calculate the electron Green functions since the
approximate operator behaves
differently from the original one in the physical Hilbert space. Nevertheless, $\langle z_\alpha\rangle
= \langle \tilde{z}_\alpha\rangle$. With Eq.~\eqref{Eq:Zapprox}, Eq.~\eqref{Eq:HSsinglesite} is then
approximated to be
\begin{eqnarray}
 \label{HStilde} H^{\mathrm{mf}}_S &\approx& \sum_{\alpha\beta} \epsilon^{\alpha\beta}
 \left(\langle \tilde{z}^\dagger_{\alpha}\rangle \tilde{z}_{\beta} + \langle \tilde{z}_{\beta}\rangle
 \tilde{z}^\dagger_{\alpha} \right) \nonumber\\
&& + \sum_\alpha \left(\frac{\lambda_{\alpha}}{2} + \bar{\epsilon}_\alpha
\eta_\alpha\right) (\hat{n}^a_{\alpha} - \hat{n}^b_{\alpha}) + H^S_{\mathrm{int}},
\end{eqnarray}
where $\bar{\epsilon}_\alpha = \sum_{\beta} (\epsilon^{\alpha\beta}\langle\tilde{z}^\dagger_\alpha\rangle
\langle\tilde{z}_\beta\rangle + \mathrm{c.c.})$.
Further using the constraint Eq.~\eqref{Eq:constraint}, we can move the term proportional to
$\eta_\alpha$ to $H^{\mathrm{mf}}_f$ by introducing an effective on-site potential $\tilde{\mu}_\alpha=2\bar{\epsilon}^{\alpha}\eta_\alpha$.
The resulting mean-field Hamiltonians are then
\begin{eqnarray}
 \label{Eq:Hfmf}  H^{\mathrm{mf}}_f &=&  \sum_{k\alpha\beta}\left[ \epsilon^{\alpha\beta}_{k} \langle \tilde{z}^\dagger_\alpha \rangle \langle \tilde{z}_\beta \rangle + \delta_{\alpha\beta}(\Delta_\alpha-\lambda_\alpha+\tilde{\mu}_\alpha-\mu)\right] f^\dagger_{k\alpha} f_{k\beta},\nonumber\\
 \\
 \label{Eq:HSmf}  H^{\mathrm{mf}}_S &=& \sum_{\alpha\beta} \left[ \epsilon^{\alpha\beta} \left(\langle \tilde{z}^\dagger_{\alpha}\rangle \tilde{z}_{\beta} + \langle \tilde{z}_{\beta}\rangle \tilde{z}^\dagger_{\alpha} \right)
+ \delta_{\alpha\beta} \frac{\lambda_{\alpha}}{2} (\hat{n}^a_{\alpha} - \hat{n}^b_{\alpha}) \right] \nonumber\\
&& + H^S_{\mathrm{int}},
\end{eqnarray}
where $\epsilon^{\alpha\beta}_{k}=\frac{1}{N}\sum_{ij} t^{\alpha\beta}_{ij} e^{ik(r_i-r_j)}$, and $\delta_{\alpha\beta}$ is Kronecker's delta function. Eq.~\eqref{Eq:Hfmf} and Eq.~\eqref{Eq:HSmf} are the main formulation of our slave-spin mean-field theory. The mean-field parameters $\langle \tilde{z}_\alpha\rangle$ and $\lambda_\alpha$ can then be solved self-consistently.
In the non-interacting limit, it
is easy to check that $Z_\alpha=|\langle \tilde{z}_\alpha\rangle|^2=1$
can be achieved by taking $\lambda_\alpha=\tilde{\mu}_\alpha$; the quasiparticle weights are equal to $1$ as it should be.

\subsection{Comparison with the $Z_2$ slave-spin theory}
Here we compare our $U(1)$ slave-spin theory with the $Z_2$ slave-spin theory.
One advantage of the $U(1)$ slave-spin theory over the $Z_2$ slave-spin approach is that it can
be directly generalized to the multi-orbital systems with non-zero crystal field splitting and/or away from half-filling.
As an example, we consider a two-orbital model at half-filling and with equal bandwidth but with
a finite crystal field splitting. In the non-interacting limit, we expect the quasiparticle spectral weight
$Z_{\alpha=1,2}=1$ and the spinon bandstructure is identical to the tight-binding dispersion of the $d$ electrons.
The $Z_2$ slave-spin mean-field theory failed to obtain these results. This is easy to see: Both of the two orbitals
are away from half-filling in the presence of the crystal field splitting. According to Eq.~\eqref{Eq:HSSmf},
to obtain $Z_{1(2)}=|\langle O_{1(2)}\rangle|^2=1$, it is necessary that $\lambda_1=-\lambda_2\neq0$.
In absence of the potential $\tilde{\mu}$ in Eq.~\eqref{Eq:Hfmf}, this already distorts the bandstructure of the spinons from the original tight-binding form. Hence one can not obtain the desired spinon filling as required by the constraint. However, in our theory, $Z_\alpha=1$ is guaranteed by the condition $\lambda_\alpha=\tilde{\mu}_{\alpha}$. These two potentials cancel out exactly as seen in Eq.~\eqref{Eq:Hfmf}. Therefore, the spinon bandstructure is identical to the tight-binding one, and the non-interacting limit is properly recovered.

However, we find that at the mean-field level and in the symmetry broken phases of the bosons/spins,
the two theories have very similar forms. To see this explicitly, we compare the mean-field Hamiltonians
of the two theories in the slave-spin representation. In this representation, the mean-field Hamiltonian
of the $U(1)$ slave-spin theory can be obtained by performing a Schwinger-boson-to-spin mapping
to the Hamiltonian $H^{\mathrm{mf}}_S$ in Eq.~\eqref{Eq:HSmf}. $H^{\mathrm{mf}}_S$ then reads
\begin{eqnarray}
 \label{Eq:HSSmf} H^{\mathrm{mf}}_{S} &=& \sum_{\alpha\beta} \left[\epsilon^{\alpha\beta}
 \left(\langle O^\dagger_\alpha\rangle O_\beta+ \langle O_\beta\rangle O^\dagger_\alpha\right)
 + \delta_{\alpha\beta}\lambda_\alpha S^z_\alpha\right] \nonumber\\
 &&+ H_{\mathrm{int}}(\mathbf{S}),
\end{eqnarray}
where
\begin{equation}\label{Eq:OU1}
O^\dagger_\alpha=\langle P^+_\alpha\rangle S^+_\alpha \langle P^-_\alpha\rangle,
\end{equation}
$P^\pm_\alpha=\sqrt{1/2+\delta\pm S^z_\alpha}$, and $Z_\alpha=|\langle O_\alpha\rangle|^2$.
Surprisingly, we see that the mean-field Hamiltonian of the $Z_2$ theory takes exactly the same form
as in Eq.~\eqref{Eq:HSSmf} if define
\begin{equation}\label{Eq:OZ2}
O_\alpha^\dagger=(\langle P^-_\alpha\rangle\langle P^+_\alpha\rangle-1) S^-_\alpha +S^+_\alpha.
\end{equation}
Interestingly, the two definitions in Eq.~\eqref{Eq:OU1} and Eq.~\eqref{Eq:OZ2} gives the same
quasiparticle spectral weight $Z_\alpha=|\langle O_\alpha\rangle|^2$.
In models with a single orbital or degenerate multiple orbitals, $\tilde{\mu}_\alpha$ in Eq.~\eqref{Eq:Hfmf}
becomes orbital independent and can thus be absorbed into the chemical potential.
Therefore, in these cases, in the metallic phase, the $U(1)$ slave-spin mean-field theory and the $Z_2$ theory
gives the same results (up to a constant in the free energies).
It should be stressed that, even for single-orbital or degenerate multi-orbital models, the agreement between the two theories is obtained only in the ordered phase of the slave spins.

Generally, the constructions in the two formulations are different in the sense already mentioned.
In the $U(1)$ slave-spin theory, the operator equation Eq.~\eqref{Eq:SSCreate} has a $U(1)$ gauge redundancy,
and allows a proper Mott insulator phase.

\section{Mott Transition in a Five-Orbital Model for Iron Pnictides}\label{Sec:MottTransition}
\begin{figure}
 \begin{center}
  \includegraphics[width=80mm]{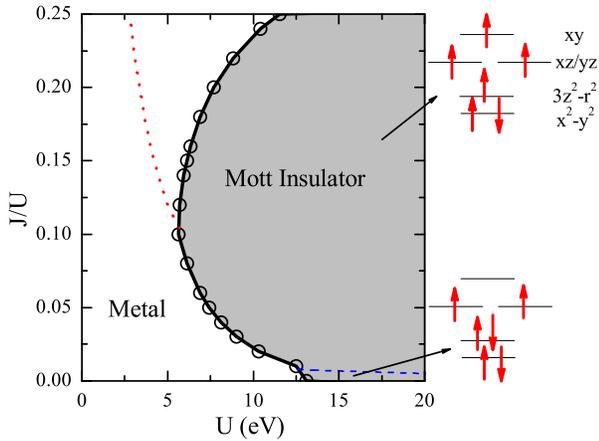}
 \end{center}
\caption{Ground-state phase diagram of the five-orbital model at $n=6.0$. The solid curve with symbols
shows the boundary between the metallic and Mott insulating phases. The dotted line shows a crossover
in the metallic phase where the system changes from a weakly correlated metal to a strongly correlated metal.
The dashed line indicates a low-spin to high-spin transition in the Mott insulating phase.
The atomic configurations corresponding to the low-spin and high-spin
Mott states are illustrated on the right side.}\label{Fig:1}
\end{figure}

\begin{figure}
 \begin{center}
  \includegraphics[width=80mm]{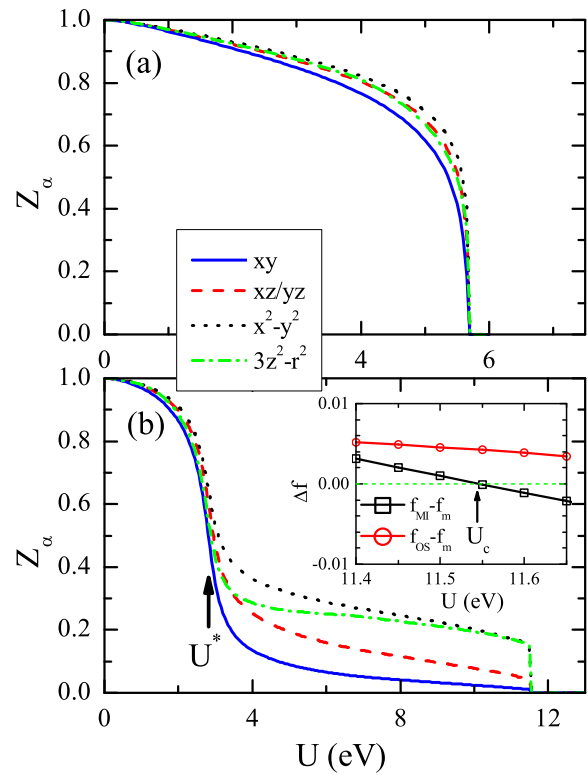}
 \end{center}
\caption{(a): Evolution of the quasiparticle spectral weight with $U$ at $n=6.0$ and $J/U=0.1$.
(b): Same as (a) but at $J/U=0.25$. The inset shows the difference of free energies ($\Delta f$)
between three competing states for the same model parameters, with $f_{\mathrm{m}}$, $f_{\mathrm{MI}}$,
and $f_{\mathrm{OS}}$ respectively denote the free energies of the metallic, Mott insulating,
and orbital selective Mott states.}\label{Fig:2}
\end{figure}

In a previous paper~\cite{Yu11,footnote} we have discussed the metal-to-insulator transition
in two- and four-orbital models. For both models we find a metal-to-Mott-insulator transition at finite $U$.
We also find that the nature of the phase transition and the critical value $U_c$ can
be strongly influenced by the Hund's rule coupling. These results are based on the assumption
that the system under study is at half-filling. But the parent iron pnictides are away from half-filling,
with six $d$ electrons occupying five orbitals in each iron atom.
To consider this more involved case,
here we use the $U(1)$ slave-spin mean-field theory to
study the five-orbital Hubbard model for iron pnicitdes.
We take the tight-binding parameters as those proposed in Ref.~\onlinecite{Graser09} for
the parent LaOFeAs. The interaction part of the Hamiltonian is as given
in Eq.~\eqref{Eq:Ham_int}. For simplicity, here we consider only the density-density interactions and
drop the spin-flip and pair-hopping terms in Eq.~\eqref{Eq:Ham_int}. The results with the full interactions
are qualitatively similar~\cite{Yu11}.

Fig.~\ref{Fig:1} shows the ground-state phase diagram at electron filling $n=6$ of the five-orbital model
in the $J$-$U$ plane. We find a single transition from a metal to an insulator. At $J=0$, $U_c\approx13.1$ eV.
In this case, we have also independently determined $U^{SR}_c\approx11.0$ eV from the slave-rotor
mean-field theory~\cite{Tal}. For a model with $M$-fold-degenerate orbitals,
the ratio $U_c/U^{SR}_c=(M+1)/M$.~\cite{FlorensGeorges,deMedici05}
This relation approximately holds in the five-orbital model, in which we find $U_c/U^{SR}_c\approx1.19$.
It can be understood by the fact that in this model, the largest crystal field splitting ($\Delta\approx0.5$ eV) is
relatively small compared to the full bandwidth ($D\approx4.0$ eV), so the orbitals are nearly degenerate.

A non-zero Hund's rule coupling strongly affects the MIT and the nature of both the insulating
and metallic states. In the insulating phase, we find that the degenerate $xz$ and $yz$ orbitals
are always at half filling, and hence in a Mott insulating state; while the $x^2-y^2$ orbital is fully occupied,
in a band insulating state. The transition is therefore an orbital selective MIT.~\cite{Yu11}
Due to the interplay of $J$ and crystal field splitting $\Delta$, the other two orbitals can be either
in a Mott insulating ($J>\Delta$), or in a band insulating state ($J<\Delta$). Accordingly,
the five-orbital model can accommodate either a high-spin ($S=2$) or a low-spin ($S=1$) Mott state,
as illustrated in Fig.~\ref{Fig:1}. These two states are separated by a low-spin-to-high-spin
transition inside the insulating phase.

In the models at half filling, $U_c$ decreases monotonically with increasing $J/U$. But the phase
diagram of the five-orbital model shows a significant difference: as $J/U$ is increased
from zero, $U_c$
first decreases for $J/U\lesssim0.1$,
but then increases with $J/U$ for $J/U\gtrsim 0.1$.
We can understand this by estimating and comparing the Mott gaps in the four- and five-orbital
models in two limiting cases: $0<J\ll \Delta$ and $J\gg \Delta$. The Mott gap $G^M$ measures
the distance between the upper and lower Hubbard bands in a Mott insulator.
It can be approximated by $G^M\approx G^A-D$;
here $G^A=E^{n+1}+E^{n-1}-2E^{n}$ is the excitation gap in the atomic limit, and $E^{n}$ is
the energy of the atomic configuration with $n$ electrons.
A reasonable estimate of $U_c$ can be obtained from
$G^A\sim D$. For $0<J\ll \Delta$, the configuration
for the undoped compound is the $S=1$ low-spin state. For this configuration, $G^A=U+J$ in both the four- and five-orbital models. Hence when $J/U$ is small, in both models we expect $U_c\sim D/(1+J/U)$, decreasing with $J/U$. But for $J\gg \Delta$, the configuration is the $S=2$ high-spin state, and $G^A$ depends on the electron filling of the high-spin state. In the four-orbital model, the system is at half-filling, $G^A\sim U+3J$, and $U_c\sim D/(1+3J/U)$, still decreasing with $J/U$. But in the five-orbital model with $n=6$, $G^A\sim U-3J$. This gives $U_c\sim D/(1-3J/U)$, which increases with $J/U$. Therefore, we expect a non-monotonic behavior of $U_c$ with increasing $J/U$, which is seen in the numerical results shown in Fig.~\ref{Fig:1}.

The Hund's rule coupling also affects the properties in the metallic state. In Fig.~\ref{Fig:2},
we compare the evolution of quasiparticle spectral weight $Z_\alpha$ for the same model but at two different $J/U$ values.
In both cases, the insulating phase is the $S=2$ high-spin Mott state.
On the other hand, $Z_\alpha$ behaves very differently in the metallic phases. At $J/U=0.1$, $U_c\approx D$,
 and $Z_\alpha$ drops rapidly down to zero only at $U\approx U_c$.
 The orbital dependence of $Z_\alpha$ is weak. At $J/U=0.25$, the Mott transition
 takes place at $U\approx 3D$. But $Z_\alpha$ drops rapidly to small but non-zero values at $U\approx2.7$ eV.
 This rapid drop allows us to defines a crossover scale $U^*$ in the metallic state.
 In Fig.~\ref{Fig:1}, we plot this crossover line in the phase diagram. For large $J/U$, $U^*$ can be smaller than $D$.
 $U^*$ increases with decreasing $J/U$ and the crossover line ends when it crosses the MIT phase boundary
 at $J/U\approx0.11$.
 At $U<U^*$ the spectral properties of the system is similar to its non-interacting limit,
 with weakly renormalized quasiparticle spectral weights. But for $U>U^*$, the quasiparticle spectral
 weights are strongly suppressed. In this regime, electron correlations are sufficiently
 strong in the metallic phase even at $U\lesssim D$, and the system is close to a Mott insulator.
 So $U^*$ roughly separates the regimes of a weakly correlated metal and a strongly correlated metal.
 The strongly correlated metallic phase has the features prescribed in the incipient Mott picture.
 As another remarkable observation, we find that for $U>U^*$, $Z_\alpha$ become
 significantly orbital dependent.
 For example, at $J/U=0.25$ and for $U\gtrsim 4$ eV, $Z_{xy}\lesssim0.1$ and
 is much smaller than that of the other orbitals. This implies that the system is close
 to an orbital selective Mott phase (OSMP). We have calculated the groundstate energy
 of an OSMP with $xy$ orbital insulating but all other orbitals metallic, and compare it with the
 energies of the metallic and insulating solutions in the inset of Fig.~\ref{Fig:2}.
 Though the OSMP never becomes the true ground state, it is indeed energetically close.
 The stabilization of an OSMP requires a high-spin configuration~\cite{deMedici09}.
 This is consistent with the observation that a threshold is needed to trigger the strongly correlated metallic state.

\section{Conclusion}\label{Sec:Conclusions}
We have developed a $U(1)$ slave-spin theory, which allows the study of
 metal-to-Mott-insulator transition in both single- and multi-orbital systems.
 For models with a single orbital or multiple degenerate orbitals,
 we show that the mean-field theory in the slave-spin-condensed phase is mathematically
 equivalent to that of the previous $Z_2$ slave-spin mean-field theory.
For
 models with multiple non-degenerate orbitals, our $U(1)$ slave-spin formulation provides
 proper descriptions for both the metal and Mott-insulating phases.

We have applied the $U(1)$ slave-spin approach to study
a five-orbital Hubbard model for the parent iron pnictides. We find that the model exhibits a metal-to-Mott-insulator
transition. The interplay between the Hund's rule coupling and crystal field splittings strongly affect
the Mott transition and the associated phases. The insulating phase can be either an $S=1$ low-spin
Mott state or an $S=2$ high-spin Mott state, depending on the strength of the Hund's rule coupling.
In the metallic phase, a crossover between a weakly correlated to a strongly correlated metallic phase
exists when the Hund's rule coupling is beyond a threshold. Inside the strongly correlated metallic phase,
the quasiparticle spectral weights are strongly suppressed, in agreement with the incipient Mott picture.
In this phase and in the vicinity of the Mott transition, we find that an orbital selective Mott phase
has ground state energies which are nearly as competitive as those of the metallic and Mott insulating states.

This work has been supported in part by
NSF Grant No. DMR-1006985 and the Robert A. Welch Foundation
Grant No. C-1411.


\begin{thebibliography}{}
\bibitem{deMedici05} L. de'Medici, A. Georges, and S. Biermann, Phys. Rev. B {\bf 72}, 205124 (2005).

\bibitem{Kamihara_FeAs} Y. Kamihara, T. Watanabe, H. Hirano, and H. Hosono, J. Am. Chem. Soc. {\bf 130}, 3296 (2008).

\bibitem{Zhao_CPL08}
Z. A. Ren {\it et al.}, Chin. Phys. Lett. {\bf 25}, 2215 (2008).

\bibitem{Rotter08}
M. Rotter, M. Tegel, and D. Johrendt,
Phys. Rev. Lett. {\bf 101}, 107006 (2008).

%
\bibitem{Hsuetal08} F.-C. Hsu {\it et al.},
Proc. Natl. Acad. Sci. {\bf 105}, 14262 (2008).

\bibitem{JGuo10}
J. Guo {\it et al.}, Phys. Rev. B \textbf{82}, 180520(R) (2010).

\bibitem{MFang11} M. Fang {\it et al.},
EPL {\bf 94}, 27009 (2011).

\bibitem{delaCruz} C. de la Cruz {\it et al.}, Nature (London) {\bf 453}, 899 (2008).

\bibitem{Dongetal08} J. Dong, {\it et al.}, Europhys. Lett. {\bf 83}, 27006 (2008).

\bibitem{Qazilbash09} M. M. Qazilbash {\it et al.}, Nat. Phys. {\bf 5}, 647 (2009).

\bibitem{Huetal09} W. Z. Hu {\it et al.}, Phys. Rev. Lett. {\bf 101}, 257005 (2008).

\bibitem{SiNatPhys09} Q. Si, Nat. Phys. {\bf 5}, 629 (2009).

\bibitem{ZhaoNatPhys09} J. Zhao {\it et al.}, Nat. Phys. {\bf 5}, 555 (2009).


\bibitem{Yietal09} M. Yi {\it et al.}, Phys. Rev. B {\bf 80}, 024515 (2009).

\bibitem{SiAbrahams08} Q. Si and E. Abrahams, Phys. Rev. Lett. {\bf 101}, 076401 (2008).

\bibitem{SiNJP09} Q. Si, E. Abrahams, J. Dai, and J.-X. Zhu, New J. Phys. {\bf 11}, 045001 (2009).

\bibitem{Haule}
K. Haule, J. H. Shim, and G. Kotliar
Phys. Rev. Lett. \textbf{100},
226402 (2008).

\bibitem{Kutepov10} A. Kutepov, K. Haule, S. Y. Savrasov, and G. Kotliar, Phys. Rev. B {\bf 82}, 045105 (2010).

\bibitem{Yildirim}
T. Yildirim,
Phys. Rev. Lett.
\textbf{101}, 057010 (2008).

\bibitem{Ma}
F. Ma, Z-Y Lu, and T. Xiang,
Phys. Rev. B\textbf{78},
224517 (2008).

\bibitem{Fang:08}
C. Fang {\it et al}, Phys. Rev. B, \textbf{77}, 224509 (2008).

\bibitem{Xu:08}
C. Xu, M. Muller, and S. Sachdev, Phys. Rev. B, \textbf{78}, 020501(R) (2008).

\bibitem{Daietal09} J. Dai, Q. Si, J.-X. Zhu, and E. Abrahams, Proc. Natl. Acad. Sci. {\bf 106}, 4118 (2009).

\bibitem{Uhrig}
G. S. Uhrig {\it et al.}, Phys. Rev. B, \textbf{79}, 092416 (2009).

\bibitem{KSeo} K. Seo, B. A. Bernevig,
and J.  Hu,
Phys. Rev. Lett. \textbf{101}, 206404 (2008).

\bibitem{WChen} W.-Q. Chen, K.-Y. Yang,
Y. Zhou, and F.-C. Zhang,
Phys. Rev. Lett. \textbf{102}, 047006 (2009).

\bibitem{PGoswami} P. Goswami, P. Nikolic, and Q. Si,
EPL \textbf{91}, 37006 (2010).

\bibitem{Moreo} A. Moreo, M. Daghofer,
J. A. Riera, and E. Dagotto,
Phys. Rev. B \textbf{79}, 134502 (2009).

\bibitem{Berg} E. Berg,
S. A. Kivelson, and D. J. Scalapino,
New J. Phys. \textbf{11}, 085007 (2009).

\bibitem{Lipscombe11} O. J. Lipscombe {\it et al.}, Phys. Rev. Lett. {\bf 106}, 057004 (2011).

\bibitem{Bao122A} W. Bao {\it et al.}, Chin. Phys. Lett. {\bf 28}, 086104 (2011).

\bibitem{Bao122B} W. Bao {\it et al.}, e-print arXiv:1102.3674.

\bibitem{Ye122} F. Ye {\it et al.}, Phys. Rev. Lett. {\bf 107}, 137003 (2011).

\bibitem{MaFeTe} F. Ma, W. Ji, J. Hu, Z.-Y. Lu, and T. Xiang, Phys. Rev. Lett. {\bf 102}, 177003 (2009).

\bibitem{HuJ1J2J3} C. Fang, B. A. Bernevig, and J. Hu, Europhys. Lett. {\bf 86}, 67005 (2009).

\bibitem{CaoDai11} C. Cao and J. Dai, Phys. Rev. Lett. 107, 056401 (2011).

\bibitem{YuKFeSeMag} R. Yu, P. Goswami, and Q. Si, Phys. Rev. B {\bf 84}, 094451 (2011).

\bibitem{Zhuetal10} J.-X. Zhu {\it et al.}, Phys. Rev. Lett. {\bf 104}, 216405 (2010).

\bibitem{YuVacOrder11} R. Yu, J.-X. Zhu, and Q. Si, Phys. Rev. Lett. {\bf 106}, 186401 (2011).

\bibitem{Zhou11} Y. Zhou, D.-H. Xu, F.-C. Zhang, and W.-Q. Chen, Europhys. Lett. {\bf 95}, 17003 (2011).

\bibitem{Yu11} R. Yu and Q. Si, Phys. Rev. B {\bf 84}, 235115 (2011).

\bibitem{KotliarRuckenstein} G. Kotliar and A. E. Ruckenstein, Phys. Rev. Lett. {\bf 57}, 1362 (1986).

\bibitem{FlorensGeorges} S. Florens and A. Georges, Phys. Rev. B {\bf 70}, 035114 (2004).

\bibitem{deMedici11} L. de'Medici, Phys. Rev. B {\bf 83}, 205112 (2011).

\bibitem{NandkishoreSenthil12} R. Nandkishore and T. Senthil, arXiv:1201.5998.

\bibitem{Sigrist} In this paper we refer as $Z_2$ slave-spin theory the formulation initially proposed in Ref.~\onlinecite{deMedici05} and
further developed in Ref.~\onlinecite{HassandeMedici10}. There is some difference between this approach and the one
proposed in A. R\"{u}egg, S. D. Huber, and M. Sigrist, Phys. Rev. B {\bf 81}, 155118 (2010), though both formulations involve
 a $Z_2$ invariance.

\bibitem{Castellani78} C. Castellani, C. R. Natoli, and J. Ranninger, Phys. Rev. B {\bf 18}, 4945 (1978).

\bibitem{HassandeMedici10} S. R. Hassan and L. de'Medici, Phys. Rev. B {\bf 81}, 035106 (2010).

\bibitem{footnote} In our earlier paper Ref.~\onlinecite{Yu11}, the two- and four-orbital models are already studied using the $U(1)$ slave-spin mean-field method for the five-orbital model, though the method itself was not explicictly introduced there.

\bibitem{Graser09} S. Graser, T. A. Maier, P. J. Hirschfeld, and D. J. Scalapino, New J. Phys. {\bf 11}, 025016 (2009).

\bibitem{Tal} T. Einav, R. Yu, and Q. Si, unpublished (2012).

\bibitem{deMedici09} L. de'Medici, S. R. Hassan, M. Capone, and X. Dai, Phys. Rev. Lett. {\bf 102}, 126401 (2009).
\end{thebibliography}
\end{document}